\documentstyle[11pt,amssymb,graphics]{article}

\begin{document}

\begin{center}
{\large {\bf Perturbative evolution of far off-resonance driven two-level
systems: Coherent population trapping, localization, and harmonic 
generation
\vspace*{0.6cm}\\}} {V. Delgado\footnote{e-mail: vdelgado@ull.es}
and J. M. Gomez Llorente\footnote{e-mail: jmgomez@ull.es}
{\vspace*{.2cm}}}\\{\it Departamento de F\'\i sica Fundamental II,\\%
Universidad de La Laguna, 38205-La Laguna, Tenerife, Spain
\vspace*{0.7cm}\\}
\end{center}

\begin{abstract}
The time evolution of driven two-level systems in the far off-resonance
regime is studied analytically. We obtain a general first-order perturbative
expression for the time-dependent density operator which is applicable
regardless of the coupling strength value. In the strong field 
regime, our perturbative expansion remains valid even when the far 
off-resonance condition is not fulfilled. We find that, in the absence of 
dissipation, driven two-level systems exhibit coherent population trapping 
in a certain region of parameter space, a property which, in the particular 
case of a symmetric double-well potential, implies the well-known 
localization of the system in one of the two wells. Finally, we show how 
the high-order harmonic generation that this kind of systems display can 
be obtained as a straightforward application of our formulation.
\end{abstract}

\hspace{6. pt} PACS number(s): 42.50.-p, 42.50.Ct

\vspace{1.4 cm}

\begin{center}
{\large {\bf {I. INTRODUCTION\vspace{.1 cm}\\}}}
\end{center}

Numerical investigations in driven symmetric bistable systems have revealed
a number of striking quantum phenomena [\ref{Grif1}]. On one hand, Grossmann 
{\it et al.} [\ref{Gros1}] have shown that the parameters of the external
driving field can be appropriately tuned so as to produce coherent
suppression of tunneling, a property that can be used to localize the
quantum system in one of the two wells. By properly choosing the shape of
the driving laser pulse, localization can be achieved even if the system is
initially in a delocalized eigenstate: the driving field can take it into a
localized state and then keep it there [\ref{Met1}]. On the other hand, this
system has also been shown to exhibit high-order harmonic generation in a
region of parameter space which overlaps to a certain extent with that where
localization occurs [\ref{Met2}]. Harmonic generation is a consequence of
the fact that, in a strong field, the induced dipole moment responds with
frequencies that are integer multiples of the laser frequency, thus giving 
rise to the appearance of the corresponding peaks in the emission spectrum.

An interesting aspect of driven double-well systems is that many of their
relevant features can be captured in a simple two-level model. Using this
kind of approach it has been shown [\ref{Jgom1}] that, in a wide parameter
range [\ref{Gros2}], the localization conditions can be correctly obtained
as a zeroth-order result of a perturbative analysis in the small parameter
$\Delta _0/\omega _{{\rm {L}}}$, with $\Delta _0$ being the
transition frequency of the two-level system and $\omega _{{\rm {L}}}$ the
driving field frequency. Specifically, in this far off-resonance regime
($\Delta _0/\omega _{{\rm {L}}}\ll 1$), localization was found to occur 
at the zeros of the Bessel function $J_0\left(2\Omega _0/\omega _{{\rm {L}}}
\right) $, where $\Omega _0$ is the Rabi frequency. 
In contrast, first order perturbation theory is required at least to
account for the high-order harmonic generation that occurs in the strong 
field regime of driven two-level systems [\ref{Sund1}]. 
In this respect, several approaches have
been developed, primarily aimed at obtaining a perturbative first-order
solution for the equation of motion governing the evolution of the induced
dipole moment [\ref{Yu1}--\ref{Dak1}]. All these approaches have to deal
with secular terms appearing at first order. These terms become divergent in
the long time limit and, consequently, have to be carefully resummed in 
order for the perturbative solution to be applicable at any time.

A somewhat different approach, specifically designed for the strong field
limit, has been followed in Ref. [\ref{Fras1}], where a first-order
perturbative solution was obtained by using a series expansion dual to the
Dyson series, and by resumming the corresponding secular terms by
renormalization group methods [\ref{Fras2}].

Since localization can occur in a two-level system even when the strong
field condition is not satisfied, it is interesting to have a perturbative
solution applicable regardless of the coupling strength value. Such 
a solution would permit to treat within a unified formulation the 
localization and harmonic generation properties exhibited by this kind of 
systems. 
On the other hand, and
because of the extensive use of two-level systems as a first approximation to
treat more complex physical systems, analytical solutions of this type become
of particular interest.

In this paper we derive a general first-order analytical solution for the
perturbative evolution of a driven two-level system in the far off-resonance
regime. Specifically, a first-order expression for the time-dependent
density operator of the system, which is applicable regardless of the 
coupling strength value, is obtained. Remarkably, in the strong field 
limit, our perturbative expansion turns out to be valid even away from the 
far off-resonance condition, and therefore, in this particular case, it 
becomes indistinguishable from that of Ref. [\ref{Fras1}]. Moreover, our
zeroth-order Hamiltonian already includes all the slowly varying
contributions thus preventing, unlike previous approaches, secular terms
from appearing in the corresponding first-order results.

In Sec. II we develop the formalism and obtain the dynamical evolution of
the corresponding density operator. Then, in Secs. III and IV this
formulation is applied, respectively, to the study of localization and
high-order harmonic generation. We find that, in the absence of dissipation,
driven two-level systems exhibit coherent population trapping in the far
off-resonance regime. In the particular case of a symmetric double-well
potential such a property implies the well-known localization of the system
in one of the two wells. Finally, the main conclusions are summarized in 
Sec. V.

\vspace{1.4 cm}

\begin{center}
{\large {\bf {II. TIME-DEPENDENT DENSITY OPERATOR\vspace{.1 cm}\\}}}
\end{center}

We consider a two-level system driven by a linearly polarized laser field of
frequency $\omega _{{\rm {L}}}$ and amplitude ${\bf E}_0$. The energy
difference between the upper level state $|2\rangle $ and the lower level
state $|1\rangle $ is denoted $\Delta _0$. In the dipole approximation the
Hamiltonian reads (in atomic units, $\hbar =1$)

\begin{equation}
\label{ec1}H_d=\frac{\Delta _0}2\left( \sigma _{22}-\sigma _{11}\right)
-\Omega_0\,g\left( t\right) \cos \left( \omega _{{\rm {L}}}t\right) 
\left(\sigma _{12}+\sigma _{21}\right) ,
\end{equation}
where $\sigma _{ij}\equiv |i\rangle \langle j|$ is the transition operator
and $\Omega _0\equiv E_0\mu $ is the Rabi frequency, with $\mu $ being the
(real) dipole matrix element between $|1\rangle $ and $|2\rangle $. It has
been assumed that the laser polarization vector and the dipole moment are
oriented along the same direction. Moreover, we have included in Eq. (\ref
{ec1}) an envelope function $g\left( t\right) $ in order to account for slow
variations of the intensity, as occurs at the turn on and off of the laser
pulse.

Appropriate scaling of the time-dependent Schr\"odinger equation permits 
the identification of the relevant dimensionless parameters. 
In this
respect, especially suitable for our purposes is the variable change $\tau
=\omega _{{\rm {L}}}t$, which yields the following dimensionless Hamiltonian

\begin{equation}
\label{ec1b}H=\frac{\Delta _0}{2\omega _{{\rm {L}}}}\left( \sigma
_{22}-\sigma _{11}\right) -\frac{\Omega _0}{\omega _{{\rm {L}}}}f\left( \tau
\right) \cos \left( \tau \right) \left( \sigma _{12}+\sigma _{21}\right) , 
\end{equation}
where $f\left( \tau \right) \equiv g(\tau /\omega _{{\rm {L}}})$.

On the other hand, particularly convenient for studying the localization
properties associated with the above Hamiltonian are the following coherent
superposition states

\begin{equation}
\label{ec2}|{\rm r}\rangle =\frac 1{\sqrt{2}}\left( |1\rangle +|2\rangle
\right) , 
\end{equation}

\begin{equation}
\label{ec3}|{\rm l}\rangle =\frac 1{\sqrt{2}}\left( |1\rangle -|2\rangle
\right) , 
\end{equation}
which, in the case of a symmetric double-well potential, correspond to
states localized on the right and left wells, respectively. In the basis $%
\left\{ |{\rm r}\rangle ,|{\rm l}\rangle \right\} $ the Hamiltonian (\ref
{ec1b}) clearly displays the symmetry of the system. Indeed, in this basis
Eq. (\ref{ec1b}) takes the form

\begin{equation}
\label{ec4}H=-\frac{\Delta _0}{2\omega _{{\rm {L}}}}\left( \sigma _{{\rm lr}%
}+\sigma _{{\rm rl}}\right) -\frac{\Omega _0}{\omega _{{\rm {L}}}}f\left(
\tau \right) \cos \left( \tau \right) \left( \sigma _{{\rm rr}}-\sigma _{%
{\rm ll}}\right) , 
\end{equation}
which, for a driving field with constant amplitude ($f\left( \tau 
\right) =1$), is manifestly symmetric under the combined transformation 
${\rm r}\longleftrightarrow {\rm l}$ and $\tau \rightarrow \tau +\pi $.

As already said in the Introduction, in the far off-resonance regime, i.e.,
when $\Delta _0/\omega _{{\rm {L}}}\ll 1$, localization occurs (for $f\left(
\tau \right) =1$) at the zeros of the Bessel function $J_0\left( 2\Omega
_0/\omega _{{\rm {L}}}\right) $. This is the regime in which we will be
primarily interested. Thus, $\Delta _0/\omega _{{\rm {L}}}$ is going to be,
in principle, the small parameter of our perturbative analysis, which, 
consequently, will be applicable regardless of the value of the coupling 
strength $\Omega _0/\omega _{{\rm {L}}}$.
That is, in the far off-resonance regime our perturbative series will 
remain valid in both the strong and the weak field limit.

In order to obtain a small Hamiltonian, proportional to $\Delta _0/\omega _{%
{\rm {L}}}$, we perform the following unitary transformation

\begin{equation}
\label{ec5}U(\tau )=e^{-i\phi (\tau )\left( \sigma _{{\rm rr}}-\sigma _{{\rm %
ll}}\right) }, 
\end{equation}
with

\begin{equation}
\label{ec6}\phi (\tau )=\!\int \!d\tau \frac{\Omega _0}{\omega _{{\rm {L}}}}%
f\left( \tau \right) \cos \left( \tau \right) \simeq \frac{\Omega _0}{\omega
_{{\rm {L}}}}f\left( \tau \right) \sin \left( \tau \right) ,
\end{equation}
where, in the last step, we have used the fact that the pulse envelope 
$f\left( \tau \right) $ is a very slowly varying function within a laser
period. The transformed Hamiltonian now takes the form

\begin{equation}
\label{ec8}H^{\prime }=UHU^{+}-iU\dot{U}^{+}=-\frac{\Delta _0}{2\omega_{{\rm 
{L}}}}\left( e^{2i\phi (\tau )}\sigma_{{\rm lr}}+e^{-2i\phi (\tau )}
\sigma _{{\rm rl}}\right) , 
\end{equation}
where $\dot{U}^{+}$ denotes the derivative of the adjoint of $U$ with respect 
to the dimensionless time $\tau $.

In order to prevent secular terms from appearing at first order in the
perturbative expansion, it is necessary to incorporate all the slowly varying
terms in the zeroth-order Hamiltonian. With this purpose, and using again
the fact that $f\left( \tau \right) $ hardly changes in a laser period, we
express the time-dependent coefficients in Eq. (\ref{ec8}) as the Fourier
series

\begin{equation}
\label{ec11}e^{\pm 2i\phi (\tau )}=\sum_{n=-\infty }^{+\infty }J_n\left[ \pm 
\frac{2\Omega _0}{\omega _{{\rm {L}}}}f\left( \tau \right) \right] e^{in\tau
}\equiv \Lambda _0+\Lambda _{\pm }(\tau ),
\end{equation}
where

\begin{equation}
\label{ec12}\Lambda _0\equiv J_0\left[ \frac{2\Omega _0}{\omega _{{\rm {L}}}}%
f\left( \tau \right) \right] ,
\end{equation}

\begin{equation}
\label{ec13}\Lambda _{\pm }(\tau )\equiv \sum_{n=1}^{+\infty }(\pm
1)^nJ_n\left[ \frac{2\Omega _0}{\omega _{{\rm {L}}}}f\left( \tau \right)
\right] \left( e^{in\tau }+(-1)^ne^{-in\tau }\right) .
\end{equation}
Substituting Eq. (\ref{ec11}) into Eq. (\ref{ec8}) we then arrive at

\begin{equation}
\label{ec14}H^{\prime }=H_0^{\prime }+\Delta H^{\prime }, 
\end{equation}
with

\begin{equation}
\label{ec14a}H_0^{\prime }\equiv -\frac{\Delta _0}{2\omega _{{\rm {L}}}}%
\Lambda _0\left( \sigma _{{\rm lr}}+\sigma _{{\rm rl}}\right) , 
\end{equation}

\begin{equation}
\label{ec14b}\Delta H^{\prime }\equiv -\frac{\Delta _0}{2\omega _{{\rm {L}}}}%
\left( \Lambda _{+}(\tau )\sigma _{{\rm lr}}+\Lambda _{-}(\tau )\sigma _{%
{\rm rl}}\right) .
\end{equation}
The slowly varying part $H_0^{\prime }$ is going to be considered our
zeroth-order Hamiltonian while, in the far off-resonance regime, $\Delta
H^{\prime }$ becomes a small perturbation. Incidentally, note that in the
strong field limit, that is, when

\begin{equation}
\label{ec15}\zeta \equiv \frac{2\Omega _0}{\omega _{{\rm {L}}}}f\left( \tau
\right) \gg 1,\;\;\;\;{\rm \ for\;}{\em any\;}\tau 
\end{equation}
the Bessel functions entering Eq. (\ref{ec13}) become, for 
$n \lesssim n_{{\rm c}}\sim \zeta $ [\ref{Abram}],

\begin{equation}
\label{ec16}J_n\left( \zeta \right) \approx \sqrt{\frac 2{\pi \zeta }}\cos
\left( \zeta -\frac{n\pi }2-\frac \pi 4\right) , 
\end{equation}
while, for $n\gtrsim n_{{\rm c}}$ they decay very fast as $J_n\left( \zeta
\right) \sim \left( e\zeta /2n\right) ^n/\sqrt{2\pi n}$. Hence, the
perturbation $\Delta H^{\prime }$ takes now the form

\begin{equation}
\label{ec17}\Delta H^{\prime }=-\frac{\Delta _0}{2\sqrt{\omega _{{\rm {L}}%
}\Omega _0f\left( \tau \right) }}\left( \Pi _{+}(\tau )\sigma _{{\rm lr}%
}+\Pi _{-}(\tau )\sigma _{{\rm rl}}\right) , 
\end{equation}
with

\begin{equation}
\label{ec18}\Pi _{\pm }(\tau )\approx (1/\sqrt{\pi })\sum_{n=1}^{n_{{\rm c}%
}}(\pm 1)^n\left( e^{in\tau }+(-1)^ne^{-in\tau }\right) \cos \left( \zeta -%
\frac{n\pi }2-\frac \pi 4\right) . 
\end{equation}
Thus, in the strong field limit, $\Delta H^{\prime }$ turns out to be
proportional to the parameter $\Delta _0/\sqrt{\omega _{{\rm {L}}}\Omega
_0f\left( \tau \right) }$, which has the interesting consequence that
whenever the latter becomes small (for {\em any} $\tau $) our perturbative
results remain valid irrespective of the value of 
$\Delta _0/\omega _{{\rm {L}}} $.

We now proceed to solve perturbatively the quantum Liouville equation for
the transformed density operator $\rho ^{\prime }$, which contains all the
dynamical information about the system,

\begin{equation}
\label{ec19}\frac{\partial \rho ^{\prime }}{\partial \tau }=-i\left[
H^{\prime },\rho ^{\prime }\right] ={\cal L}^{\prime }\rho ^{\prime }=\left( 
{\cal L}_0^{\prime }+\Delta {\cal L}^{\prime }\right) \rho ^{\prime }, 
\end{equation}

\begin{equation}
\label{ec19b}\rho ^{\prime }(\tau )=U(\tau )\rho (\tau )U^{+}(\tau ). 
\end{equation}

The linear operators ${\cal L}_0^{\prime }$ and $\Delta {\cal L}^{\prime }$
entering Eq. (\ref{ec19}) represent the quantum Liouville operators
corresponding to the Hamiltonians $H_0^{\prime }$ and $\Delta H^{\prime }$,
respectively. This evolution equation can be exactly solved to the lowest
order. In doing so, one finds the following expressions for the matrix
elements of the density operator in the states $|1\rangle $ and $|2\rangle $,

\begin{equation}
\label{ec20}\rho _{ii}^{(0)\prime }(\tau )=\rho _{ii}^{\prime
}(0),\;\;\;\;\;\;\;i=1,2 
\end{equation}

\begin{equation}
\label{ec21}\rho _{12}^{(0)\prime }(\tau )=\rho _{12}^{\prime }(0)e^{i\frac{%
\Delta _0\Lambda _0}{\omega _{{\rm {L}}}}\tau }.
\end{equation}

It is worth noting that according to Eqs. (\ref{ec19b}) and (\ref
{ec20}), to the lowest order, the populations of the states $U^{+}(\tau
)|1\rangle $ and $U^{+}(\tau )|2\rangle $ remain constant. This is a direct
consequence of the fact that, for $f\left( \tau \right) \approx {\rm cte}$,
the states 
\begin{equation}
\label{ec23a}U^{+}(\tau )|1\rangle =\cos \phi (\tau )|1\rangle +i\sin \phi
(\tau )|2\rangle ,
\end{equation}
\begin{equation}
\label{ec23b}U^{+}(\tau )|2\rangle =\sin \phi (\tau )|1\rangle -i\cos \phi
(\tau )|2\rangle ,
\end{equation}
become the zeroth-order Floquet states of the system corresponding
to the quasienergies $-\Delta _0\Lambda _0/2$ and 
$+\Delta _0\Lambda _0/2$, respectively.

The zeroth-order density operator $\rho ^{(0)\prime }$ can be written as

\begin{equation}
\label{ec24}\rho ^{(0)\prime }(\tau )=e^{{\cal L}_0^{\prime }\tau }\rho
^{\prime }(0)=\sum_{i=1,2}\rho _{ii}^{\prime }(0)|i\rangle \langle i|+\left(
\rho _{12}^{\prime }(0)e^{i\frac{\Delta _0\Lambda _0}{\omega _{{\rm {L}}}}%
\tau }|1\rangle \langle 2|+{\rm h.c.}\right) .
\end{equation}
Notice that
this expression provides, in turn, an useful operational definition for 
the evolution operator $e^{{\cal L}_0^{\prime }\tau }$. 

With the aim of deriving a first-order perturbative expression for the
time-dependent density operator, we rewrite the quantum Liouville 
equation (\ref{ec19}) in the interaction representation with respect to 
${\cal L}_0^{\prime }$. Specifically,

\begin{equation}
\label{ec27}\frac{\partial \rho _{{\rm I}}^{\prime }}{\partial \tau }=\Delta 
{\cal L}_{{\rm I}}^{\prime }\,\rho _{{\rm I}}^{\prime }(\tau ), 
\end{equation}
where

\begin{equation}
\label{ec28a}\rho _{{\rm I}}^{\prime }=e^{-{\cal L}_0^{\prime }\tau }\rho
^{\prime }(\tau ), 
\end{equation}

\begin{equation}
\label{ec28b}\Delta {\cal L}_{{\rm I}}^{\prime }=e^{-{\cal L}_0^{\prime
}\tau }\Delta {\cal L}^{\prime }e^{{\cal L}_0^{\prime }\tau }. 
\end{equation}

The solution of Eq. (\ref{ec27}) can be expressed as the following infinite
series

\begin{equation}
\label{ec29}\rho _{{\rm I}}^{\prime }(\tau )=\rho _{{\rm I}}^{\prime
}(0)+\int_0^\tau d\tau _1\Delta {\cal L}_{{\rm I}}^{\prime }(\tau _1)\rho _{%
{\rm I}}^{\prime }(0)+\int_0^\tau d\tau _1\int_0^{\tau _1}d\tau _2\Delta 
{\cal L}_{{\rm I}}^{\prime }(\tau _1)\Delta {\cal L}_{{\rm I}}^{\prime
}(\tau _2)\rho _{{\rm I}}^{\prime }(0)+\ldots  
\end{equation}
This is a perturbative expansion in the small parameter $\Delta _0/\omega _{%
{\rm {L}}}$ which, in the strong field limit, becomes an expansion in the
small parameter $\Delta _0/\sqrt{\omega _{{\rm {L}}}\Omega _0f\left( \tau
\right) }$, as already said. Thus, up to first order, one obtains

\begin{equation}
\label{ec30}\rho ^{\prime }(\tau )=e^{{\cal L}_0^{\prime }\tau }\rho
^{\prime }(0)+e^{{\cal L}_0^{\prime }\tau }\int_0^\tau d\tau _1e^{-{\cal L}%
_0^{\prime }\tau _1}\Delta {\cal L}^{\prime }(\tau _1)e^{{\cal L}_0^{\prime
}\tau _1}\rho ^{\prime }(0)+O\left( \epsilon ^2\right) ,
\end{equation}
where $\epsilon $ represents, in general, the small parameter characterizing
the expansion. On the other hand, the integrand on the right hand side of
Eq. (\ref{ec30}) takes the form

\begin{eqnarray}
e^{{\cal L}_0^{\prime }(\tau -\tau _1)}\Delta {\cal L}^{\prime }(\tau _1)e^{%
{\cal L}_0^{\prime }\tau _1}\rho ^{\prime }(0)=-ie^{{\cal L}_0^{\prime
}(\tau -\tau _1)}\left[ \Delta H^{\prime }(\tau _1),e^{{\cal L}_0^{\prime
}\tau _1}\rho ^{\prime }(0)\right] \nonumber \\
\label{ec31}=\sum_{i=1,2}\xi _{ii}^{\prime }(\tau _1)|i\rangle \langle
i|+\left( \xi _{12}^{\prime }(\tau _1)e^{i\frac{\Delta _0\Lambda _0}{\omega
_{{\rm {L}}}}(\tau -\tau _1)}|1\rangle \langle 2|+{\rm h.c.}\right) ,
\end{eqnarray}
where

\begin{equation}
\label{ec32}\xi _{jk}^{\prime }(\tau )\equiv -i\langle j|\left[ \Delta
H^{\prime }(\tau ),e^{{\cal L}_0^{\prime }\tau }\rho ^{\prime }(0)\right]
|k\rangle ,
\end{equation}
and the definition of the propagator $e^{{\cal L}_0^{\prime }(\tau -\tau _1)}
$, as provided by Eq. (\ref{ec24}), has been used. Substituting then Eq. (%
\ref{ec24}) into Eq. (\ref{ec32}) one obtains, after some lengthy algebra,

\begin{equation}
\label{ec33a}\xi _{jj}^{\prime }(\tau )\equiv (-1)^j\frac{\Delta _0}{\omega
_{{\rm {L}}}}\sum_{n=0}^{+\infty }J_{2n+1}\left( \zeta \right) \sin \left[
(2n+1)\tau \right] \left( \rho _{12}^{\prime }(0)e^{i\frac{\Delta _0\Lambda
_0}{\omega _{{\rm {L}}}}\tau }+{\rm h.c.}\right) ,
\end{equation}

\begin{eqnarray}
\xi _{12}^{\prime }(\tau )&\equiv& -\frac{\Delta _0}{\omega_{{\rm 
{L}}}}\sum_{n=0}^{+\infty }J_{2n+1}\left( \zeta \right) \sin 
\left[ (2n+1)\tau \right] \left[ \rho _{22}^{\prime }(0)-
\rho _{11}^{\prime }(0)\right] 
\nonumber \\
\label{ec33b}& & \mbox{}+2i\frac{\Delta _0}{\omega _{{\rm {L}}}}
\sum_{n=1}^{+\infty}J_{2n}\left( \zeta \right) \cos \left( 2n\tau \right) 
\rho _{12}^{\prime}(0)e^{i\frac{\Delta _0\Lambda _0}{\omega_{{\rm 
{L}}}}\tau }, 
\end{eqnarray}

\begin{equation}
\label{ec33c}\xi _{21}^{\prime }(\tau )=\xi _{12}^{\prime *}(\tau ), 
\end{equation}
with $\zeta = {2\Omega _0}f\left( \tau\right)/{\omega _{{\rm {L}}}}$.
Substitution of the above
expressions into Eq. (\ref{ec31}) and then into Eq. (\ref{ec30}) leads to
the following final result

\begin{eqnarray}
\rho ^{\prime }(\tau )=\left[ \rho _{11}^{\prime }(0)+\frac{\Delta _0}{%
\omega _{{\rm {L}}}}\alpha (\tau )\right] |1\rangle \langle 1|+\left[ \rho
_{22}^{\prime }(0)-\frac{\Delta _0}{\omega _{{\rm {L}}}}\alpha (\tau
)\right] |2\rangle \langle 2| \nonumber \\
\label{ec34}+\left\{ \left[ \rho _{12}^{\prime }(0)+\frac{\Delta _0}{\omega
_{{\rm {L}}}}\beta (\tau )\right] e^{i\frac{\Delta _0\Lambda _0}{\omega _{%
{\rm {L}}}}\tau }|1\rangle \langle 2|+{\rm h.c.}\right\} +O\left( \epsilon
^2\right) ,
\end{eqnarray}
where

\begin{equation}
\label{ec35a}\alpha (\tau )\equiv \frac 12\sum_{n=0}^{+\infty }\frac{%
J_{2n+1}\left( \zeta \right) }{2n+1}\left\{ \rho _{12}^{\prime }(0)\left(
e^{i\left[ (2n+1)+\frac{\Delta _0\Lambda _0}{\omega _{{\rm {L}}}}\right]
\tau }+e^{-i\left[ (2n+1)-\frac{\Delta _0\Lambda _0}{\omega _{{\rm {L}}}}%
\right] \tau }-2\right) +{\rm h.c.}\right\} , 
\end{equation}

\begin{eqnarray}
\beta (\tau )&\equiv& \frac 12 \sum_{n=0}^{+\infty }
\frac{J_{2n+1}\left( \zeta
\right) }{2n+1}\left[ \rho _{22}^{\prime }(0) - \rho_{11}^{\prime }(0)
\right]\left\{ e^{i\left[ (2n+1)-\frac{\Delta _0\Lambda _0}
{\omega _{{\rm {L}}}}
\right] \tau }+e^{-i\left[ (2n+1)+\frac{\Delta _0\Lambda _0}{\omega _{{\rm 
{L}}}}\right] \tau }-2\right\} \nonumber \\
\label{ec35b}
& & \mbox{} + 2i\sum_{n=1}^{+\infty }\frac{J_{2n}\left( \zeta 
\right) }{2n}
\rho _{12}^{\prime }(0)\sin \left( 2n\tau \right) . 
\end{eqnarray}
Thus, the first-order density operator of the driven two-level system is
finally given by

\begin{equation}
\label{ec36}\rho (\tau )=U^{+}(\tau )\rho ^{\prime }(\tau )U(\tau ),
\end{equation}
with $U(\tau )$ defined in Eqs. (\ref{ec5}) and (\ref{ec6}). This
expression, which contains no secular terms, is our central result and is
applicable not only in the far off-resonance limit, i.e.,

\begin{equation}
\label{ec37}\Delta _0/\omega _{{\rm {L}}}\ll 1, 
\end{equation}
but also in the regime where

\begin{equation}
\label{ec26b}\Delta _0/\omega _{{\rm {L}}}\ll \sqrt{\Omega _0f\left( \tau
\right) /\omega _{{\rm {L}}}}\gg 1. 
\end{equation}

\vspace{1.4 cm}

\begin{center}
{\large {\bf {III. COHERENT POPULATION TRAPPING AND LOCALIZATION%
\vspace{.1 cm}\\}}}
\end{center}

Coherent population trapping and localization properties are conveniently
analyzed in terms of the diagonal matrix elements of $\rho (\tau )$ in the
basis $\left\{ |{\rm r}\rangle ,|{\rm l}\rangle \right\} $. By using Eqs. (%
\ref{ec2}) and (\ref{ec36}) one can express the population $%
\langle {\rm r}|\rho (\tau )|{\rm r}\rangle $ as

\begin{equation}
\label{ec25}\rho _{{\rm rr}}(\tau )=\frac 12\left\{ 1+\left[ \rho
_{12}^{\prime }(\tau )+\rho _{21}^{\prime }(\tau )\right] \right\} .
\end{equation}
Taking into account that $\rho ^{\prime }(0)=\rho (0)$ one obtains, after
substitution of Eq. (\ref{ec21}) into Eq. (\ref{ec25}), the following
zeroth-order expression

\begin{equation}
\label{ec26}\rho _{{\rm rr}}^{(0)}(\tau )=\frac 12+|\rho _{12}(0)|\cos
\left( \frac{\Delta _0\Lambda _0}{\omega _{{\rm {L}}}}\tau +\varphi \right) ,
\end{equation}
where $\rho _{12}(0)=|\rho _{12}(0)|e^{i\varphi }$. Equation (\ref{ec26})
shows that, in the absence of dissipation, the populations of the $|{\rm r}%
\rangle $ and $|{\rm l}\rangle $ states, which in general oscillate with a
frequency $\Delta _0\Lambda _0$ (curves (a) and (b) in Fig. 1),
become trapped whenever
the parameters of the driving field are so tuned that the Bessel function 
$\Lambda _0\equiv J_0\left( 2\Omega _0f\left( \tau \right) /
\omega _{{\rm {L}}}\right) $ attains a zero (curve (c) in Fig. 1). 
Remarkably, such a population trapping occurs
regardless of the system initial state. This is a general zeroth-order 
result that remains valid as long as conditions (\ref{ec37}) or 
(\ref{ec26b}) are satisfied. 
In the particular case of a symmetric double-well potential, 
$|{\rm r}\rangle $ and $|{\rm l}\rangle $ become states localized on the 
right and left wells, respectively, so that, when the system is initially 
in either of the two wells, one obtains the peculiar localization property
reported in previous works [\ref{Gros1}--\ref{Gros2}].

\begin{figure}
{\par\centering \resizebox*{12cm}{!}{\rotatebox{-90}
{\includegraphics{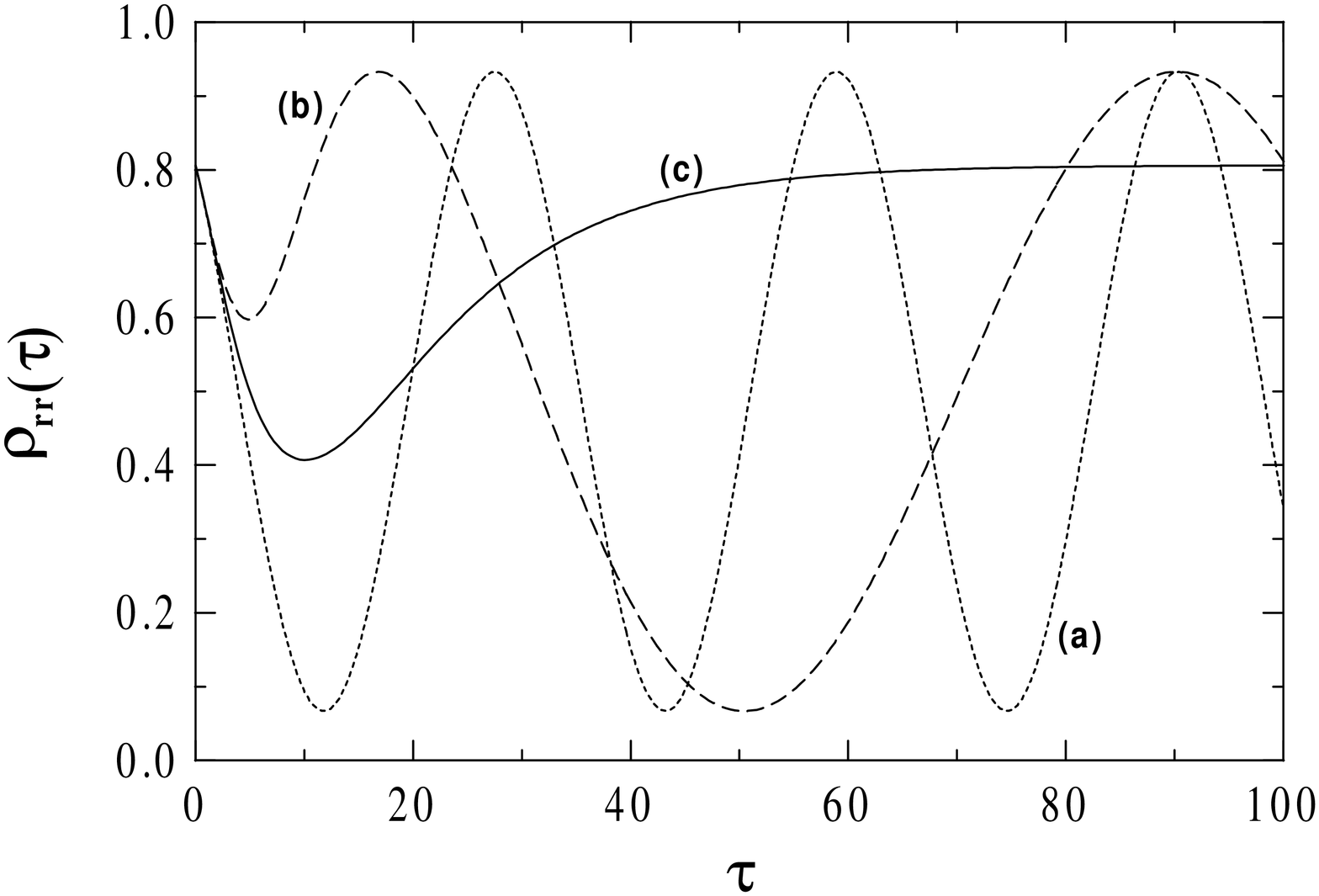}}} \par}

\caption{\small 
Zeroth-order population $\rho _{\rm{rr}}(\tau )$
for a pulse envelope $f(\tau )=1-\exp (- \tau /10)$ and $\Delta _{0}/
\omega _{\rm{L}}=0.2$, $|\rho _{12}(0)|=\sqrt{3}/4$, $\varphi =\pi /4$;
(a) $\Omega _{0}/\omega _{\rm{L}}=0$; (b) $\Omega _{0}/
\omega _{\rm{L}}=1.8$; (c) $\Omega _{0}/\omega _{\rm{L}}=1.202$.}
\end{figure}

Substituting Eq. (\ref{ec34}) into Eq. (\ref{ec25}) one finds the 
following final expression for the first-order time-dependent population

\begin{eqnarray}
\rho _{{\rm rr}}(\tau )&=&\frac 12 + |\rho _{12}(0)|\cos 
\left( \frac{\Delta
_0\Lambda _0}{\omega _{{\rm {L}}}}\tau +\varphi \right) \nonumber \\
& & \mbox{} + \frac{\Delta _0}{\omega _{{\rm {L}}}}
\left[ \rho _{22}(0)-\rho
_{11}(0)\right] \sum_{n=0}^{+\infty }\frac{J_{2n+1}\left( \zeta \right) }
{2n+1}\left( \cos (2n+1)\tau -\cos \frac{\Delta _0\Lambda _0}
{\omega _{{\rm {L}}}}\tau \right) \nonumber \\
\label{ec38}& & \mbox{} -2\frac{\Delta _0}{\omega _{{\rm {L}}}}|\rho
_{12}(0)|\sum_{n=1}^{+\infty }\frac{J_{2n}\left( \zeta \right) }{2n}\sin
\left( 2n\tau \right) \sin \left( \frac{\Delta _0\Lambda _0}{\omega _{{\rm {L%
}}}}\tau +\varphi \right) +O\left( \epsilon ^2\right) .
\end{eqnarray}

Equation (\ref{ec38}) shows that now the populations evolve undergoing
rapidly oscillating changes of the order of $\epsilon $, superimposed to the
much slower dominant oscillating evolution of frequency $\Delta _0
\Lambda_0$ (see curve (a) in Fig. 2). 
In general, these high frequency oscillations
cannot be eliminated by the external field and as a consequence, unlike the
previous case, to this order it is not possible to achieve exact coherent
trapping for any system initial state. Indeed, when $\Lambda _0=0$ the
population $\rho _{{\rm rr}}(\tau )$ evolves describing small-amplitude
rapid oscillations (with frequencies that are integer multiples of the 
laser frequency) about the initial population $\rho _{{\rm rr}}(0)$
[curve (b) in Fig. 2]. 
Yet, exact first-order coherent trapping can still be obtained by properly 
choosing the initial state. Taking $\rho _{11}(0)=\rho _{22}(0)=1/2$ and 
$\varphi =0$, Eq. (\ref{ec38}) becomes, when $\Lambda _0=0$,

\begin{equation}
\label{ec39}\rho _{{\rm rr}}(\tau )=\frac 12+\rho _{12}(0)=\rho _{{\rm rr}%
}(0).
\end{equation}
\begin{figure}
{\par\centering \resizebox*{12cm}{!}{\rotatebox{-90}
{\includegraphics{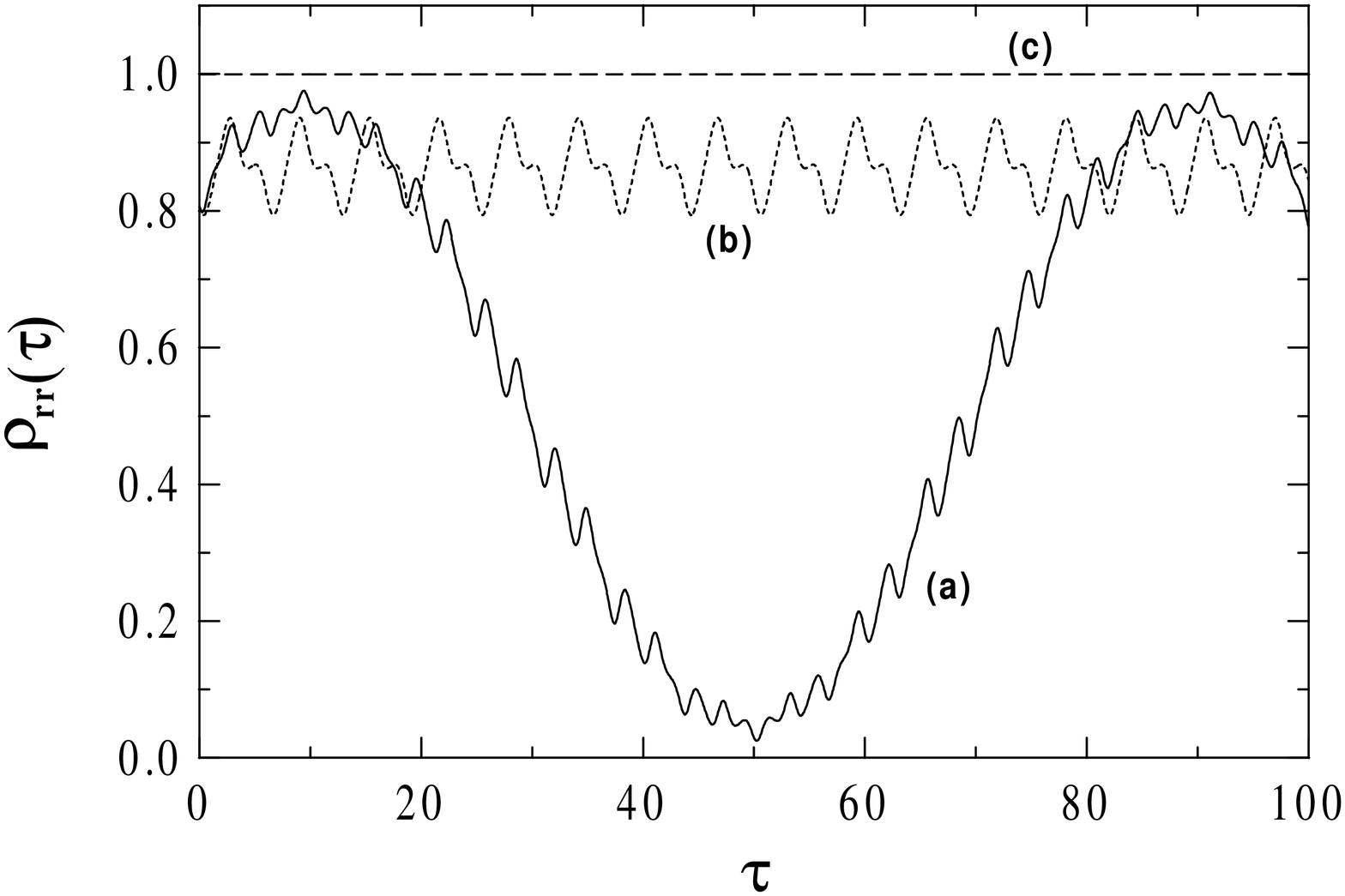}}} \par}

\caption{\small
First-order population $\rho _{{\rm rr}}(\tau )$ for $f(\tau )=1$
and $\Delta _0/\omega _{{\rm {L}}}=0.2$; (a) $\rho _{11}(0)=3/4$, 
$\rho_{22}(0)=1/4$, $|\rho _{12}(0)|=\sqrt{3}/4$, $\varphi =\pi /4$, 
$\Omega_0/\omega _{{\rm {L}}}=1.8$; (b) $\rho _{11}(0)=3/4$, 
$\rho _{22}(0)=1/4$, $|\rho _{12}(0)|=\sqrt{3}/4$, $\varphi =\pi /4$, 
$\Omega _0/\omega _{{\rm {L}}}=1.202$; 
(c) $\rho _{11}(0)=\rho _{22}(0)=1/2$, $|\rho _{12}(0)|=1/2$, 
$\varphi =0$, $\Omega _0/\omega _{{\rm {L}}}=1.202$.}
\end{figure}
That is, when the system is initially in either the state $|{\rm r}\rangle $
or $|{\rm l}\rangle $, first-order coherent population trapping can be
achieved by simply choosing the parameters of the driving field in such a
way that $\Lambda _0=0$
(curve (c) in Fig. 2). 
Notice that, as before, in the special case of a
symmetric double-well potential, this result implies the localization of the
system in one of the two wells.

On the other hand, it is interesting to note that according to Eq. (\ref
{ec38}) the population of a system initially located in the right well of a
symmetric double-well potential, evolves in the limit of zero driving field as

\begin{equation}
\label{ec40}\rho _{{\rm rr}}(t)=\cos ^2\left( \frac{\Delta _0}2t\right) 
\end{equation}
so that, the well-known result for the undriven system is recovered, as it
should be.

Notice that the pulse envelope $f(t)$ simply modifies the way in which
the system reaches its steady state (see Fig. 1). 
As a consequence, localization cannot be perturbatively achieved starting 
from an initial delocalized eigenstate. 
Indeed, as is apparent from Eq. (\ref{ec38}), in this case the
system remains always delocalized, within terms of the order of $\epsilon $.
This result, which is in contradiction with the numerical results obtained
in Refs. [\ref{Met1},\ref{Met2}], simply reflects that such a localization
process is a nonperturbative effect that cannot be accounted for by our
treatment.

\vspace{1.4 cm}

\begin{center}
{\large {\bf {IV. HARMONIC GENERATION\vspace{.1 cm}\\}}}
\end{center}

Driven two-level systems has been extensively used as convenient models to
understand the basic mechanism underlying the phenomenon of high-order
harmonic generation [\ref{Sund1},\ref{Yu1},\ref{Dak1},\ref{Gaut1}--\ref
{Gaut2}]. This is so primarily because, as pointed out by Sundaram and
Milonni [\ref{Sund1}], such a simple system already exhibits the main
features observed experimentally in the emission spectrum of atoms in very
intense laser fields, namely, the existence of a plateau in the harmonic
spectrum followed by a sharp cutoff. This, in turn, is an indication of the
fact that such features are intrinsic properties of strongly driven systems.

The purpose of this Section is to show how the previous formulation leads
straightforwardly to the well-known results of high-order harmonic
generation in strongly driven two-level systems.

The coherent part of the emission spectrum $S(\omega )$ is proportional to $%
\left| d(\omega )\right| ^2$, with $d(\omega )$ being the Fourier component
at the frequency $\omega $ of the induced dipole moment $\langle d(t)\rangle 
$ [\ref{Sund1},\ref{Gaut1}]

\begin{equation}
\label{ec41}S(\omega )\propto \left| d(\omega )\right| ^2=\left| \frac
1T\int_{t_0}^{t_0+T}\!dt\,e^{i\omega t}\langle d(t)\rangle \right| ^2. 
\end{equation}
The time evolution of $\langle d(t)\rangle $ follows directly from Eq. (\ref
{ec38}). Indeed, taking into account that the quantum dipole operator is
given by

\begin{equation}
\label{ec42}d=\mu \left( \sigma _{12}+\sigma _{21}\right) =\mu \left( \sigma
_{{\rm rr}}-\sigma _{{\rm ll}}\right) , 
\end{equation}
one finds

\begin{equation}
\label{ec43}\langle d(t)\rangle ={\rm Tr}\left[ \rho (t)d\right] =\mu \left[
\rho _{{\rm rr}}(t)-\rho _{{\rm ll}}(t)\right] =2\mu \left( \rho _{{\rm rr}%
}(t)-1/2\right) ,
\end{equation}
where, in the last step, we have used that $\rho _{{\rm rr}}(t)+
\rho _{{\rm ll}}(t)=1$. Therefore, by substituting Eq. (\ref{ec38}) into 
Eq. (\ref{ec43}),
and then this latter into Eq. (\ref{ec41}) one can immediately obtain the
corresponding emission spectrum. In doing so, one finds the well-known
result that $S(\omega )$ consists, in general, of three types of peaks:

\begin{enumerate}
\item[i)]  Low frequency components located at $\omega =\pm \Delta _0\Lambda
_0$ and intensities proportional to 
\begin{equation}
\label{ec44}\left| d_{\pm \Delta _0\Lambda _0}\right| ^2=\mu ^2|\rho
_{12}(0)|^2.
\end{equation}

\item[ii)]  Hyper-Raman lines located at frequencies $2n\omega _{{\rm {L}}%
}\pm \Delta _0\Lambda _0$ and intensities proportional to 
\begin{equation}
\label{ec45}\left| d_{2n\omega _{{\rm {L}}}\pm \Delta _0\Lambda _0}\right|
^2=\left| \mu \frac{\Delta _0}{\omega _{{\rm {L}}}}|\rho _{12}(0)|\frac{%
J_{2n}\left( \zeta \right) }{2n}\right| ^2\;\;\;\;\;\;n=1,2,3,\ldots 
\end{equation}

\item[iii)]  Odd harmonic components located at frequencies$\;\left(
2n+1\right) \omega _{{\rm {L}}}$ and intensities proportional to
\end{enumerate}

\begin{equation}
\label{ec52}
\left| d_{\left( 2n+1\right) \omega _{{\rm {L}}}}\right| ^2=\left| \mu 
\frac{\Delta _0}{\omega _{{\rm {L}}}}\left[ \rho _{22}(0)-\rho _{11}(0)
\right] \frac{J_{2n+1}\left( \zeta \right) }{2n+1}\right| ^2
\;\;\;\;\;\;n=0,1,2,\ldots  
\end{equation}
Formulas of this type have been previously obtained by Ivanov and Corkum [%
\ref{Yu1}] and Dakhnovskii and Bavli [\ref{Dak1}].

It is interesting to note that exact first-order coherent population
trapping (localization) has a twofold manifestation in the system
spectroscopic properties: a zero frequency peak appears and the spectrum
consists of only even harmonic components.

On the other hand, the appearance in the strong field regime ($\zeta \gg 1$)
of the plateau and corresponding cutoff in the harmonic spectrum follows
from the asymptotic behaviour of the Bessel functions, which for 
$n\lesssim n_{{\rm c}}\sim \zeta $ are given by Eq. (\ref{ec16}) while 
for $n\gtrsim n_{{\rm c}}$ they drop very rapidly as $\left( e\zeta 
/2n\right) ^n/\sqrt{2\pi n}$.

\vspace{1.4 cm}

\begin{center}
{\large {\bf {V. CONCLUSION\vspace{.1 cm}\\}}}
\end{center}

Driven two-level systems display a number of interesting features that are
common to more general driven systems. This fact, makes them convenient
starting points for analyzing more complex physical systems. Moreover, under
certain circumstances, they allow for an analytical treatment which would be
otherwise impossible. Such analytical treatments provide detailed
information on the system response to specific changes in the external
parameters, and this information can be used to control its dynamical
evolution.

In this work we have studied analytically the time evolution of driven
two-level systems in the far off-resonance regime ($\Delta _0/\omega _{{\rm {%
L}}}\ll 1$). This was done by performing a perturbative analysis based on a
convenient zeroth-order Hamiltonian which takes care of divergent secular
terms. In this way, we obtained a general first-order expression for the
time-dependent density operator which is valid regardless of the coupling 
strength value. 
Interestingly enough, in the strong field regime, this
expression turns out to be applicable even when the far off-resonance
condition is not satisfied. Indeed, the perturbative analysis remains valid
as long as the condition $\Delta _0/\omega _{{\rm {L}}}\ll \sqrt{\Omega
_0f\left( \tau \right) /\omega _{{\rm {L}}}}\gg 1$ holds.

The analytical formulation presented in this paper makes it possible to
treat in a unified framework different aspects of the dynamical evolution of
driven two-level systems. In particular, from the time evolution of the
populations it immediately follows that the well-known phenomenon of
tunneling suppression in a symmetric double-well potential can be considered
as a specific manifestation of a more general population trapping phenomenon.

To the lowest order, regardless of the system initial state, the populations 
of the coherent superpositions $|{\rm r}\rangle $ and $|{\rm l}\rangle $ 
become trapped, 
in the absence of dissipation, whenever the parameters of the
driving field are tuned in such a way that the Bessel function $J_0\left(
2\Omega _0f\left( \tau \right) /\omega _{{\rm {L}}}\right) $ vanishes. To
first-order, however, small-amplitude rapid oscillations in the time 
evolution of the populations appear, which, in general, cannot be eliminated
by the external field. As a consequence, to this order, such a result only
remains valid in an approximate way. Nonetheless, exact first-order coherent
population trapping still occurs when the system is initially in either the
state $|{\rm r}\rangle $ or $|{\rm l}\rangle $. In the particular case of a
symmetric double-well potential, in which $|{\rm r}\rangle $ and $|{\rm l}%
\rangle $ become states localized on the right and left wells, respectively,
this result implies the localization of the system in one of the two wells.

The present formulation also leads straightforwardly to the well-known
results of high-order harmonic generation in strongly driven two-level
systems.

Of course physical systems exhibit a number of interesting features which
are of nonperturbative origin and, consequently, cannot be accounted for 
by a perturbative treatment of the type presented here. In spite of this, 
the present formulation can be useful as a starting point for analyzing a 
large variety of driven systems in the perturbative regime.

\vspace{1.2 cm}

\begin{center}
{\large {\bf {ACKNOWLEDGMENTS \vspace{.1 cm}\\}}}
\end{center}

This work has been supported by DGESIC (Spain) under Project 
No. PB97-1479-C02-01.

\newpage
\vspace{1.2 cm}

\begin{center}
{\large {\bf REFERENCES \vspace{.4cm}}}
\end{center}

\begin{enumerate}
\item  \label{Grif1} For a review see, for example, M. Grifoni and P.
H\"anggi, Phys. Rep. {\bf 304}, 229 (1998).

\item  \label{Gros1} F. Grossmann, T. Dittrich, P. Jung, and P. H\"anggi,
Phys. Rev. Lett. {\bf 67}, 516 (1991).

\item  \label{Met1} R. Bavli and H. Metiu, Phys. Rev. Lett. {\bf 69}, 1986
(1992).

\item  \label{Met2} R. Bavli and H. Metiu, Phys. Rev. A {\bf 47}, 3299
(1993).

\item  \label{Jgom1} J. M. Gomez Llorente and J. Plata, Phys. Rev. A {\bf 45}%
, R6958 (1992).

\item  \label{Gros2} F. Grossmann and P. H\"anggi, Europhys. Lett. {\bf 18},
571 (1992).

\item  \label{Sund1} B. Sundaram and P. W. Milonni, Phys. Rev. A {\bf 41},
R6571 (1990).

\item  \label{Yu1} M. Yu. Ivanov and P. B. Corkum, Phys. Rev. A {\bf 48},
580 (1993).

\item  \label{Met3} Y. Dakhnovskii and H. Metiu, Phys. Rev. A {\bf 48}, 2342
(1993).

\item  \label{Dak1}Y. Dakhnovskii and R. Bavli, Phys. Rev. B {\bf 48}, 11020
(1993).

\item  \label{Fras1} M. Frasca, Phys. Rev. A {\bf 60}, 573 (1999).

\item  \label{Fras2} M. Frasca, Phys. Rev. A {\bf 56}, 1548 (1997).

\item  \label{Abram} M. Abramowitz and I. A. Stegun, {\it Handbook of
Mathematical Functions} (Dover, New York, 1972).

\item  \label{Gaut1} F. I. Gauthey, C. H. Keitel, P. L. Knight, and A.
Maquet, Phys. Rev. A {\bf 52}, 525 (1995).

\item  \label{Pons1} M. Pons, R. Ta\"\i eb, and A. Maquet, Phys. Rev. A {\bf %
54}, 3634 (1996).

\item  \label{Gaut2} F. I. Gauthey, C. H. Keitel, P. L. Knight, and A.
Maquet, Phys. Rev. A {\bf 55}, 615 (1997).
\end{enumerate}






\end{document}